\documentclass[aoas,preprint]{imsart}
\usepackage{bm}
\RequirePackage[OT1]{fontenc}
\RequirePackage{amsthm,amsmath}
\RequirePackage[authoryear]{natbib}
\setcitestyle{authoryear,open={(},close={)}}
\RequirePackage[colorlinks,citecolor=blue,urlcolor=blue]{hyperref}
\usepackage{graphicx}
 \usepackage{multirow}
 \usepackage{lscape}
\usepackage{ulem}
\usepackage{soul}
\usepackage[figuresleft]{rotating}

\usepackage{tikz}
\usetikzlibrary{fit,positioning}
\usetikzlibrary{bayesnet}
\tikzstyle{latent} = [circle,fill=white,draw=black,inner sep=1pt,
minimum size=6pt, font=\fontsize{12}{12}\selectfont, node distance=0.5cm]
\tikzstyle{obs} = [latent,fill=gray!25]

\tikzstyle{const} = [rectangle, inner sep=2pt, node distance=0.25]

\tikzset{%
  link/.style    = { white, double = black, line width = 1.8pt,
                     double distance = 0.8pt }
}
\arxiv{arXiv:0000.0000}

\startlocaldefs
\theoremstyle{plain}

\endlocaldefs

\begin{document}

\begin{frontmatter}
\title{A Bayesian model of microbiome data for simultaneous identification of covariate associations and prediction of phenotypic outcomes}
\runtitle{Joint modeling of microbiome data }

\begin{aug}
\author{\fnms{Matthew D.} \snm{Koslovsky}\thanksref{t1}\ead[label=e1]{mkoslovsky@rice.edu}},

\author{\fnms{Kristi L.} \snm{Hoffman}\thanksref{t2}\ead[label=e2]{Kristi.Hoffman@bcm.edu}},

\author{\fnms{Carrie R.} \snm{Daniel}\thanksref{t3}\ead[label=e3]{CDaniel@mdanderson.org}},

\and
\author{\fnms{Marina} \snm{Vannucci}\thanksref{ t1}\ead[label=e1]{marina@rice.edu}}

\runauthor{M. Koslovsky et al.}

\affiliation{Rice University\thanksmark{t1}, Baylor College of Medicine\thanksmark{t2}, and \\The University of Texas MD Anderson Cancer Center\thanksmark{t3}}

\address{Department of Statistics\\ Rice University\\ Houston, TX, USA\\
\printead{e1}\\
\phantom{E-mail:mkoslovsky@rice.edu\ }}

\address{Alkek Center for Metagenomics \& Microbiome Research \\Baylor College of Medicine\\ Houston, TX, USA\\
\printead{e2}\\
}

\address{Department of Epidemiology \\ The University of Texas MD Anderson Cancer Center
\\ Houston, TX, USA\\
\printead{e3}}
\end{aug}

\begin{abstract}
One of the major research questions regarding human microbiome studies is the feasibility of designing interventions that modulate the composition of the microbiome to promote health and cure disease. This requires extensive understanding of the modulating factors of the microbiome, such as dietary intake, as well as the relation between microbial composition and phenotypic outcomes, such as body mass index (BMI). Previous efforts have modeled these data separately, employing two-step approaches that can produce biased interpretations of the results. Here, we propose a Bayesian joint model that simultaneously identifies clinical covariates associated with microbial composition data and predicts a phenotypic response using information contained in the compositional data.  Using spike-and-slab priors, our approach can handle high-dimensional compositional as well as clinical data. Additionally, we accommodate the compositional structure of the data via balances and overdispersion typically found in microbial samples. We apply our model to understand the relations between dietary intake, microbial samples, and BMI. In this analysis, we find numerous associations between microbial taxa and dietary factors that may lead to a microbiome that is generally more hospitable to the development of chronic diseases, such as obesity. Additionally, we demonstrate on simulated data how our method outperforms two-step approaches and also present a sensitivity analysis.
\end{abstract}

\begin{keyword}
\kwd{Bayesian statistics}
\kwd{joint modeling}
\kwd{multivariate count data}
\kwd{prediction}
\kwd{variable selection}
\end{keyword}

\end{frontmatter}

\section{Introduction}

Human microbiome research seeks to better understand the role of our microbial communities and how they interact with their host, respond to their environment, and influence disease \citep{xia2017hypothesis}. For example, current findings suggest that the microbiome is responsive to diet, as well as other factors, and may influence various metabolic conditions, such as obesity \citep{maruvada2017human, dao2016akkermansia,sonnenburg2016diet}. Insights into the relations between microbial composition and both endogenous and exogenous factors may help researchers design personalized intervention strategies to modulate and maintain a healthy microbiome community \citep{xu2015dietary,knights2011human}. However, complex environmental interactions with the microbiome challenge our understanding of community function and its impact on health \citep{shetty2017intestinal}.

Human microbiome studies typically have two main objectives:  (1) identifying factors that characterize the composition of the microbiome and (2) predicting biological, genetic, clinical, or experimental conditions using microbial abundance data \citep{xia2017hypothesis}. For both objectives, analysis is challenged for various reasons, including vast amounts of intra- and inter-subject heterogeneity in taxonomic abundance, as well as the compositional structure and high-dimensionality of the data. While each of these objectives have been extensively researched separately, we are unaware of any attempts to jointly model all of the data to achieve both objectives simultaneously. 

For objective (1), there are various methods available to infer relations between covariates and multivariate count data \citep{zhang2017regression}. For microbial count data, researchers have previously used Dirichlet-multinomial models, since these models can handle overdispersed data that arise from within- and between-subject variability in microbial data \citep{la2012hypothesis, chen2013variable,wadsworth2017integrative}. In exploratory research studies, researchers have used penalized likelihood approaches to simultaneously shrink unassociated covariates' regression coefficients to zero and estimate the effects of associated covariates \citep{chen2013variable,wang2017dirichlet}. The proven efficiency of these methods comes at a price, as optimization routines are challenged by complex data structures \citep{wang2017dirichlet}, and they do not fully capture the uncertainty of model selection. Alternatively, Bayesian methods are available which capture uncertainty in the model by exploring the model space using Markov chain Monte Carlo (MCMC) algorithms.  \cite{wadsworth2017integrative} recently developed a Bayesian approach for identifying Kegg orthology pathways that were associated with microbial abundance data using spike-and-slab priors for the regression coefficients. In confirmatory settings, \cite{mao2017bayesian} demonstrate how including covariates in a Bayesian graphical compositional regression model can improve accuracy in testing results and reduce false discoveries.

For objective (2), researchers may be interested in using microbial abundances to predict outcomes of interest, such as body mass index (BMI) \citep{lin2014variable,wang2017constructing}. Microbial abundance data are an example of multivariate compositional data where the magnitude of a single component depends on the sum of all the components' counts. This dependency causes inferential biases and computational challenges if the compositional data are modeled in their raw form. To properly model compositional data, log-ratio transformations are used. Various log-ratio transformations have been proposed, including additive, centered, and isometric \citep{aitchison1986statistical,egozcue2003isometric}. Isometric log-ratio transformations, in particular, allow researchers to properly model compositional data using balances to make inference on subsets of the taxa as opposed to individual taxon \citep{morton2017balance,pinto2017balances}. Balances are defined proportionally to the difference in the mean of the log-transformed abundances between two groups and are scale invariant. Thus, they can equivalently be constructed with raw counts or the relative proportion of counts. Additionally, researchers can use prior knowledge of structure in the data to construct balances \citep{morton2017balance,fivserova2011interpretation}. Once the raw compositional data are appropriately transformed, they can be used in standard analysis methods, such as linear regression and principle components analysis 
\citep{garcia2013identification,lin2014variable,hron2012linear,shi2016regression,gloor2017microbiome,silverman2017phylogenetic,mert2018compositional,pinto2017balances,chen2017multiple, bruno2016non}.

In this work, we propose a Bayesian joint modeling approach that simultaneously identifies clinical covariates associated with microbial composition data and predicts a phenotypic response using information contained in the compositional data. 
We conjecture that separate, two-step approaches may underestimate model uncertainty since the microbial composition data are typically treated as fixed when used to predict phenotypic responses. This may produce biased interpretation of the model \citep{chatfield1995model}. On the contrary, our joint modeling of all the data allows researchers to make inference on the relation between clinical measures and health outcomes, via their relation to the composition of the microbiome. Additionally, if there is a true relation between microbial composition and the phenotypic outcome, properly accommodating microbial heterogeneity based on clinical measures may result in a more accurate prediction. 
Our method is designed to accommodate high-dimensional microbial and clinical measures data, overdispersion in the count data, as well as the structure of the compositional data. 

We apply our method to understand the relation between dietary intake, taxonomic composition of the microbiome and BMI. We have available dietary assessments and oral and fecal microbiome data from an ancillary study conducted among healthy obese and lean individuals from the Houston, TX area \citep{versace2015heterogeneity}. The study was designed to assess eating behavior and the microbiome in self-reported healthy individuals. In our analysis, we find numerous associations between microbial taxa and dietary factors that may lead to a microbiome that is generally more hospitable to the development of chronic diseases, such as obesity. Additionally, we use simulated data to compare selection performance and predictive ability of our proposed method with respect to various two-step approaches, that first select covariates associated with multivariate count data and then perform variable selection on balances, constructed using estimated count probabilities for the prediction of a phenotypic outcome. 

 In section 2, we introduce our proposed joint model and describe the posterior inference. In section 3, we apply our method to data collected to investigate the relation between diet, microbial samples, and BMI.  In section 4, we perform a simulation study aimed at comparing performance with alternative approaches and present a sensitivity analysis. In section 5, we provide concluding remarks. 

\section{Methods}
Let $y_i$ be the observed phenotypic outcome for the $i^{th}$ subject, $i = 1,\dots,N$. Also, let $\boldmath{z}_i' = (z_{i,1}, \dots, z_{i,J})$ represent a $J$-dimensional vector of microbial taxa abundance counts and $x_i' = (x_{i,1},\dots, x_{i,P})$ be a vector of $P$ dietary covariates collected on the $i^{th}$ subject. In the Bayesian paradigm, inference is drawn from the posterior distribution, which is proportional to the likelihood of the observed data times the prior distribution of the parameters in the model. Here, we jointly model the compositional count and response data by parameterizing their likelihoods with a shared parameter (i.e., the probability of the compositional taxa).

In our joint modeling, we first assume that taxa counts $z_i$ follow a Multinomial distribution
\begin{equation}\label{eq:one}
z_i\sim \mbox{Multinomial}(\dot{z}_{i}|\psi_i),
\end{equation}
with $\dot{z}_{i} = \sum_{j=1}^J z_{i,j}$, and $\psi_i$ defined on the $J$-dimensional simplex  
$$S^{J-1} = \{ (\psi_{i,1}, \dots, \psi_{i,J}): \psi_{i,j} \geq 0, \forall j, \sum_{j=1}^J \psi_{i,j} = 1\}.$$

\noindent To account for overdispersion in the multivariate count data, we specify a conjugate prior on the taxa probabilities,
\begin{equation}\label{eq:two}
\psi_i\sim \mbox{Dirichlet}({\boldmath\gamma}_i),
\end{equation}
with the $J$-dimensional vector $\boldmath{\gamma_i} = (\gamma_{i,j}>0, \forall j \in J)$, similarly to \cite{wadsworth2017integrative} and \cite{la2012hypothesis}. Note that if we were only interested in identifying dietary covariates associated with the taxa count data, we could integrate out the $\psi_{i}$ and model $z_i$ with a Dirichlet-multinomial($\gamma_{i}$), similar to \cite{wadsworth2017integrative}. However for our joint model, we estimate $\psi$ since it serves as the shared parameter between the likelihood of the phenotypic response $\boldsymbol{Y}$ and compositional data $\boldsymbol{Z}$, as described below.  Next,  we incorporate dietary covariate effects into the model by using a log-linear regression framework. Specifically, we set $\lambda_{i,j} = \log(\gamma_{i,j})$ and assume 
 \begin{equation} \label{lambda}
  \lambda_{i,j} = \alpha_{j} + \sum_{p=1}^P \varphi_{jp}x_{i,p},
 \end{equation}
where $\boldmath{\varphi}_{j}  = (\varphi_{j1},\dots,\varphi_{jP})$ represents the covariates' potential relation with the $j^{th}$ compositional taxon, and 
 $\alpha_j$ is a taxon-specific intercept term. By exponentiating \eqref{lambda} we ensure positive hyperparmeters for the Dirichlet distribution. Note that while this analysis focuses on dietary factors,  other covariates, e.g., age, sex, medication use, could be included in $\boldsymbol{x}$ as well. 
 
Under this parameterization, the number of potential models to choose from when performing model selection grows quickly, even for small covariate spaces. For example, $P=10$ covariates and just $J=2$ compositional taxa results in over a million potential models. To reduce the dimension of the model we employ multivariate variable selection spike-and-slab priors  \citep{richardson2010, stingo2010} that identify dietary covariates that are associated with each compositional taxon, as opposed to spike-and-slab constructions that select variables as relevant to either all or none of the responses \citep{brown1998multivariate}. Here, we assume the covariates' inclusion in the model is characterized by a latent, $J \times P$-dimensional inclusion vector $\boldsymbol{\zeta}$. With this formulation, $\zeta_{jp}=1$ indicates that covariate $p$ is associated with compositional taxon $j$ and 0 otherwise. The prior for $\varphi_{jp}$ given $\zeta_{jp}$ follows a mixture of a normal distribution and a Dirac-delta function at zero, $\delta_{0}$, and is commonly referred to as the  spike-and-slab prior. Specifically,
\begin{equation}
\varphi_{jp}|\zeta_{jp},r_j^2 \sim \zeta_{jp} \cdot N(0,r_{j}^2) + (1-\zeta_{jp}) \cdot \delta_0(\varphi_{jp}),
\end{equation}
where $r_{j}^2$ is set large to impose a vague prior for the regression coefficients in the case of covariate inclusion.  We assume each $\zeta_{jp}$ follows a Bernoulli prior, $p(\zeta_{jp}) \sim \mbox{Bernoulli}(\omega_{jp})$,
where $\omega_{jp} \sim \mbox{Beta}(a,b)$. Integrating out $\omega_{jp}$ leads to 
\begin{equation}
p(\zeta_{jp}) = \frac{\mbox{Beta}(\zeta_{jp} + a,1 - \zeta_{jp} + b)}{\mbox{Beta}(a,b)}.
\end{equation}
Hyperparameters $a$ and $b$ can be set to impose various levels of sparsity in the model. Lastly, we assume the intercept terms $\alpha_{j}$ follow a $N(0, \sigma_{j}^2)$, where $\sigma^2_{j}$ are set large to impose vague priors.  

Next, we model the relation between the phenotypic response $\boldsymbol{Y}$ and the compositional data $\boldsymbol{Z}$ via a multivariable linear regression model. Typically, raw (or relative) compositional data used to construct balances for regression modeling are treated as fixed. In our joint model, we assume they are random and calculate balances using the compositional taxa probabilities $\boldsymbol{\psi}$. 
As such, our model is related to the broad class of methods that make distributional assumptions for covariates to reduce inferential biases  \citep{carroll2006measurement,tadesse2005bayesian,shi2016regression}.

Let the observed outcome $y_i$ be related to an $M$-dimensional set of balances following
 \begin{equation} \label{linear}
 y_i = \alpha_0 + \sum_{m = 1}^{M}\beta_m B(\psi)_{i,m}  + \epsilon_{i},
 \end{equation}

\noindent where $\alpha_0$ is an intercept term, $\beta_m$ is a regression coefficient for its respective balance as a function of $ \psi $, $B(\psi)_{i,m}$, and $\epsilon_i \sim N(0,\sigma^2)$.  Note that this formulation can easily be extended to include other covariates, in addition to the balances, that may be associated with the phenotypic response.  To demonstrate how to construct a balance, consider two non-overlapping partitions of $\psi$, $\psi_+$ and $\psi_-$. The balance calculated for this partition is defined as
\begin{equation} 
B(\psi_+,\psi_-) = \sqrt{\frac{|\psi_-||\psi_+|}{|\psi_-|+|\psi_+|}}\log \left[ \frac{g(\psi_+)}{g(\psi_-)} \right],
\end{equation}
where $|\cdot|$ is the dimension of a given subset and $g(\cdot)$ is the geometric mean defined as $\left( \prod_{r = 1}^{|\psi|}\psi_r \right)^{1/|\psi|}$. In our approach, balances are constructed using sequential binary separation \citep{egozcue2005groups}, producing $M = J-1$ potential balances in the model.
It is important to note that prediction performance of the model does not depend on the order in which the partitions are defined \citep{egozcue2005groups}. Additionally, log-ratio transformations cannot handle observed zero counts and require adjustments based on assumptions of their occurrence \citep{martin2015bayesian}. To handle zero values for the $\boldsymbol{\psi}$, we use a multiplicative replacement strategy in which zero values are replaced with relatively small pseudovalues, and the corresponding probability vector is scaled to sum to one \citep{martin2000zero}. Note that this strategy does not affect the DM portion of the model. There, zero counts are admissible. 

In practice, the dimension of the balance space can be large relative to $N$. To induce sparsity on the dimension space of the balances, we take a similar strategy as above and assume that the prior for $\beta_{m}$, conditioned upon a latent indicator $\xi_m$ and $\sigma^2$, follows
\begin{equation}
\beta_{m}|\xi_m,\sigma^2 \sim \xi_m \cdot N(0,h_{\beta}\sigma^2) + (1-\xi_m)\cdot \delta_0(\beta_m),
\end{equation}
and similarly,
\begin{equation}
p(\xi_m) = \frac{\mbox{Beta}(\xi_m + a_m, 1 - \xi_m + b_m)}{\mbox{Beta}(a_m,b_m)}.
\end{equation}
 The prior for the intercept term is $\alpha_0|\sigma^2 \sim N(0, h_{\alpha_0}\sigma^2)$.
Large values for the hyperparameters $h_{\alpha_0}$ and $h_{\beta}$ impose vague priors on the intercept term and regression coefficients, respectively. To complete the prior specification of the model, we set 
$\sigma^2 \sim \mbox{Inverse-gamma}(a_0, b_0),$ with $a_0 > 0 $ and $b_0 > 0 $. A graphical representation of our model is provided in Figure 1.  

To summarize, our joint model assumes that the distribution of the phenotypic response $\boldsymbol{Y}$ and taxa abundance counts $\boldsymbol{Z}$ are conditionally independent given the compositional taxa probabilities $\boldsymbol{\psi}$. Specifically, we assume
\begin{equation}\label{assumption}
f(\boldsymbol{Y}|\boldsymbol{\psi})f(\boldsymbol{Z}|\boldsymbol{\psi})p(\boldsymbol{\psi}|\boldsymbol{x}),
\end{equation}
where $f(\boldsymbol{Y}|\boldsymbol{\psi})$ models the prediction of the phenotypic response, based on balances calculated using the compositional taxa probabilities $\boldsymbol{\psi}$, and $f(\boldsymbol{Z}|\boldsymbol{\psi})p(\boldsymbol{\psi}|\boldsymbol{x})$ characterizes the associations between taxa abundance counts and clinical covariates. In the Supplementary Material, we provide a simulation study demonstrating the model's invariance to balance specification and how balance sparsity can improve prediction performance \citep{koslovsky2020}. 
 
\begin{sidewaysfigure}\label{graphical}
  \caption{Graphical representation of the proposed Bayesian joint model for identifying dietary intake covariates associated with microbial taxa and predicting BMI.}
  \centering
\includegraphics[scale = 0.5]{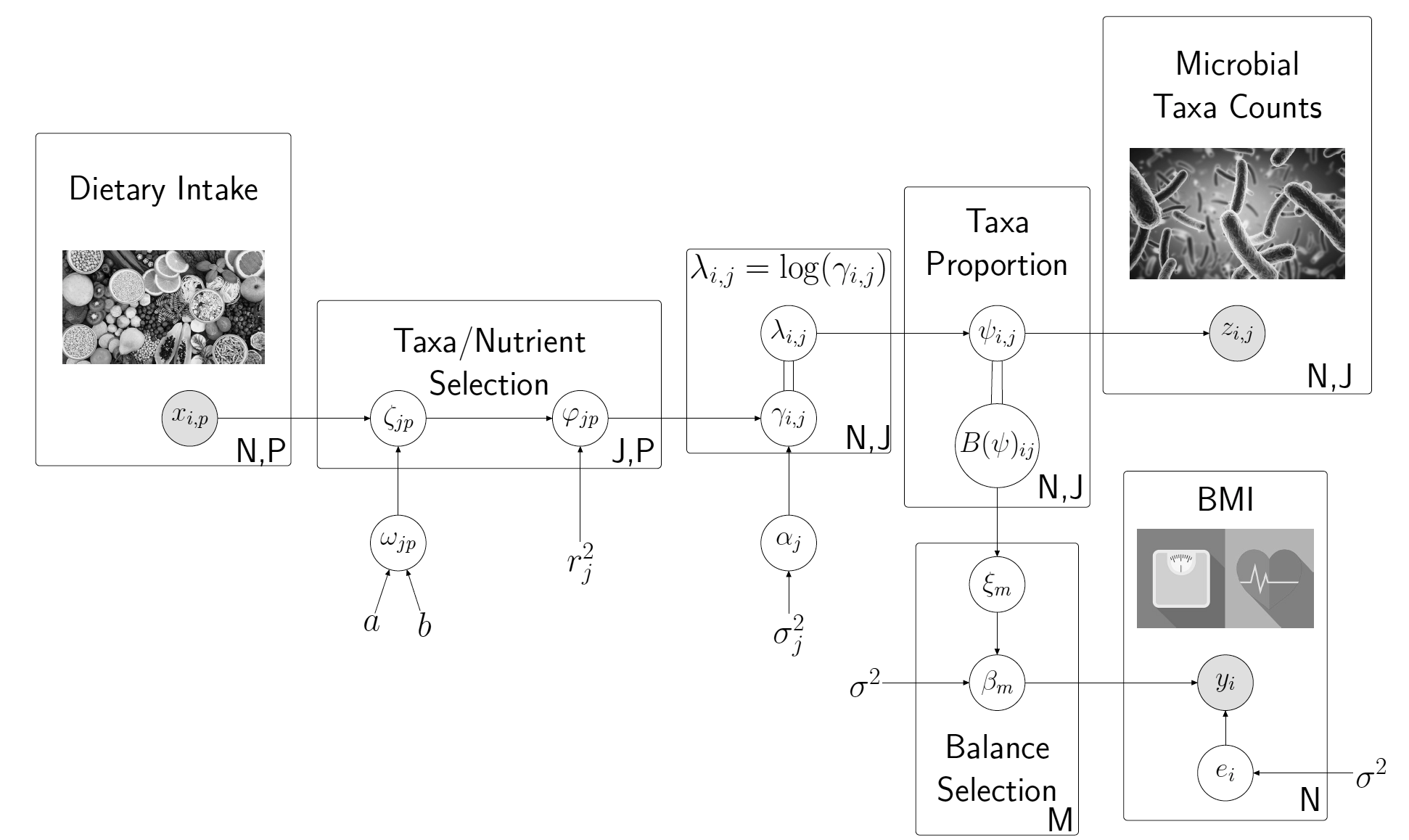}
\end{sidewaysfigure}

\subsection{Posterior Inference}
We implement a Metropolis-Hastings algorithm within a Gibbs sampler. Inspired by techniques used in Bayesian nonparametrics \citep{james2009posterior, argiento2015priori}, we adopt a data augmentation approach for the Dirichlet-multinomial portion of the model, which avoids Metropolis-Hastings updates for the taxa proportion parameters $\boldsymbol{\psi}$ and greatly aids scalability. 
First,
we integrate out $\alpha_0,\boldsymbol{\beta},$ and $\sigma^2$ in the conditional likelihood for $\boldsymbol{Y}$ to obtain a multivariate $t$-distribution,
$$\boldsymbol{Y}\sim \boldsymbol{t}_{2a_0}\left(\boldsymbol{0}_N,\frac{b_0}{a_0}(I_N + h_{\alpha}\boldsymbol{1}_N\boldsymbol{1}_N' + h_{\beta}\boldsymbol{B}(\boldsymbol{\psi})_{\boldsymbol{\xi}}\boldsymbol{B}(\boldsymbol{\psi})_{\boldsymbol{\xi}}')\right),$$
with $\boldsymbol{0}_N$ an $N-$dimensional vector of zeros, $I_N$ an $N \times N $ identity matrix, $\boldsymbol{1}_N$ an $N-$dimensional vector of ones, and $\boldsymbol{B}(\boldsymbol{\psi})_{\boldsymbol{\xi}}$ the matrix of balances included in the model. Next, we introduce latent variables $c_{i,j}$ such that $\psi_{i,j} = c_{i,j}/T_i$ with $T_i = \sum_{j=1}^Jc_{i,j}$ and reparameterize equation  \eqref{eq:one}  as
$$z_i \sim \mbox{Multinomial}(\dot{z}_i| c_i/T_i),$$
where $ c_i' = (c_{i,1},\dots,c_{i,J})$, and $c_{i,j} \sim \mbox{Gamma}(\gamma_{i,j},1)$. Then, we write the joint distribution of $z_i$ and $\psi_i$, in terms of $c_i$, as 
\begin{equation}\label{eq:joint}
p(z_i,c_i|\gamma_i)\propto \frac{c_{i,1}^{z_{i,1}} \times \cdots \times c_{i,J}^{z_{i,J}}}{T_i^{\dot{z}_i}}\prod_{j=1}^{J}\frac{1}{\Gamma(\gamma_{i,j})}c_{i,j}^{\gamma_{i,j}-1}\exp(-c_{i,j}).
\end{equation}
To avoid the calculation of the $T_i^{\dot{z}_i}$ terms, we introduce auxiliary parameters $\boldsymbol{u}' = (u_1,\dots,u_N$), such that
$u_i|T_i \sim \mbox{Gamma}(\dot{z_i},T_i)$. Using the gamma identity
$$\frac{1}{T_i^{\dot{z}_i}} = \int_0^{\infty}\frac{1}{\Gamma(\dot{z}_i)}u_i^{\dot{z}_i-1}\exp(-T_iu_i) \partial u_i ,$$
we can express \eqref{eq:joint} as 
$$p(z_i,c_i|\gamma_i)\propto \int_0^{\infty}\frac{1}{\Gamma(\dot{z}_i)}u_i^{\dot{z}_i-1}\exp(-T_iu_i) c_{i,1}^{z_{i,1}} \times \cdots \times c_{i,J}^{z_{i,J}}\prod_{j=1}^{J}\frac{1}{\Gamma(\gamma_{i,j})}c_{i,j}^{\gamma_{i,j}-1}\exp(-c_{i,j})\partial u_i. $$
Using \eqref{assumption} and transforming $\psi_i$ with $c_i$, the joint posterior distribution simplifies as proportional to 
$$f(\boldsymbol{Y}|\boldsymbol{\xi}, \boldsymbol{c})f(\boldsymbol{Z}|\boldsymbol{c})p(\boldsymbol{c}| \boldsymbol{\alpha},\boldsymbol{\varphi},\boldsymbol{\zeta}, \boldsymbol{x})p(\boldsymbol{\xi})p(\boldsymbol{\alpha})p(\boldsymbol{\varphi}|\boldsymbol{\zeta}) p(\boldsymbol{\zeta}) p(\boldsymbol{u}|\boldsymbol{c}),$$
where the integral obtained from the data augmentation technique is naturally estimated as a part of the full MCMC routine.

The generic iteration of the MCMC comprises of the following updates:

\begin{itemize}
\item Update each $\alpha_{j}$: Propose $\alpha_{j}' \sim N(\alpha_{j},0.5)$. Accept $\alpha_j'$ with probability
$$\min \left\lbrace \frac{p(\boldsymbol{c}| \boldsymbol{\alpha}',\boldsymbol{\varphi},\boldsymbol{\zeta}, \boldsymbol{x})p(\alpha'_j)}{p(\boldsymbol{c}|\boldsymbol{\alpha},\boldsymbol{\varphi},\boldsymbol{\zeta}, \boldsymbol{x})p(\alpha_j)},1  \right\rbrace .$$

\item Jointly update a $\zeta_{jp}$ and $\varphi_{jp}$ following the two-step approach proposed by \cite{savitsky2011variable}.\\
  \textit{Between-Model Step} - Randomly selecte a $\zeta_{jp}$. If $\zeta_{jp} = 1$, perform a Delete step, otherwise perform an Add step. 
\begin{itemize}
\item Delete - Propose $\zeta_{jp}' = 0$ and $\varphi_{jp}' = 0$. Accept proposal with probability 
$$\min \left\lbrace \frac{ p(\boldsymbol{c}|\boldsymbol{\alpha},\boldsymbol{\varphi}', \boldsymbol{\zeta}', \boldsymbol{x})p(\zeta'_{jp}) }{ p(\boldsymbol{c}|\boldsymbol{\alpha},\boldsymbol{\varphi}, \boldsymbol{\zeta} , \boldsymbol{x})p(\varphi_{jp}|\zeta_{jp})p(\zeta_{jp}) },1  \right\rbrace .$$
\item Add - Propose $\zeta_{jp}' = 1$. Then sample a $\varphi_{jp}' \sim N(\varphi_{jp},0.5)$.
\\ Accept proposal with probability 

$$\min \left\lbrace \frac{ p(\boldsymbol{c}|\boldsymbol{\alpha},\boldsymbol{\varphi}', \boldsymbol{\zeta}', \boldsymbol{x})p(\varphi'_{jp}|\zeta'_{jp})p(\zeta'_{jp}) }{ p(\boldsymbol{c}|\boldsymbol{\alpha},\boldsymbol{\varphi}, \boldsymbol{\zeta}, \boldsymbol{x} )p(\zeta_{jp}) },1  \right\rbrace .$$
\end{itemize}

\textit{ Within-Model Step }

\begin{itemize}
\item Propose a $\varphi_{jp}'\sim N(\varphi_{jp},0.5)$ for each covariate currently selected in the model ($\zeta_{jp}=1$). Accept each proposal with probability 
$$\min \left\lbrace \frac{ p(\boldsymbol{c}|\boldsymbol{\alpha},\boldsymbol{\varphi}',\boldsymbol{\zeta}, \boldsymbol{x})p(\varphi_{jp}'|\zeta_{jp}) }{ p(\boldsymbol{c}|\boldsymbol{\alpha},\boldsymbol{\varphi},\boldsymbol{\zeta}, \boldsymbol{x})p(\varphi_{jp}|\zeta_{jp}) },1  \right\rbrace$$

\end{itemize}

\item Update each $c_{i,j}$ via a Gibbs step:
\begin{itemize}
\item $\mbox{Gamma}(c_{i,j}|z_{i,j} + \gamma_{i,j}, u_i+1)$
\end{itemize}

\item  Update each  $u_i$ via a Gibbs step:
\begin{itemize}
\item $\mbox{Gamma}(u_i|\dot{z}_i, T_i).$
\end{itemize} 

\item Update $\xi_m$ via an Add/Delete step: Select a random $\xi_{m}$. If $\xi_{m} = 1$, perform a Delete step ($\xi_{m}' = 0$), otherwise perform an Add Step ($\xi_m' = 1$). For both Add and Delete steps, accept proposal with probability 
$$\min \left\lbrace \frac{ f(\boldsymbol{Y}|\boldsymbol{\xi}',\boldsymbol{c})p(\xi'_{m}) }{ f(\boldsymbol{Y}|\boldsymbol{\xi},\boldsymbol{c})p(\xi_{m}) }, 1  \right\rbrace .$$

\end{itemize}

For implementation, the algorithm is initiated at a set of arbitrary parameter values and then used to generate samples of the posterior distribution. After burn-in, a procedure which involves removing a subset of samples that may be influenced by initialization, the remaining samples are used for inference. To determine inclusion in the model, the marginal posterior probability of inclusion (MPPI) for each of the covariates and balances is determined by taking the average of their respective inclusion indicator's MCMC samples. Note that a covariate has a unique inclusion indicator for each of the compositional taxon. Commonly, variables are included in the model if their MPPI $\ge$ 0.50 \citep{barbieri2004optimal}. Alternatively, \cite{newton2004detecting} propose using a threshold based on a Bayesian false discovery rate to control for multiplicity. 

To evaluate the prediction accuracy of the model, cross-validation can be performed by fitting the model on a subset of the data (training set) and evaluating prediction performance on the remaining data (testing set) by calculating the prediction mean squared error. To obtain predictions of the testing outcomes, $\boldsymbol{\mathcal{Y}}$, set
\begin{equation}\label{predictY}
\hat{\boldsymbol{\mathcal{Y}}} = \hat{\alpha}_0 + \frac{1}{S}\sum_{s=1}^S \boldsymbol{B}(\ddot{\boldsymbol{\psi}})\boldsymbol{\hat{\beta}}_{\boldsymbol{\xi}^s},\end{equation}
where
$\hat{\alpha}_0 = (n + h_{\alpha_0}^{-1})^{-1}\boldsymbol{1}_n'\boldsymbol{Y}$ and
\begin{equation}
\boldsymbol{\hat{\beta}}_{\boldsymbol{\xi}^s}=\left(\boldsymbol{B}(\boldsymbol{\psi}^s)_{\boldsymbol{\xi}^s}'\boldsymbol{B}(\boldsymbol{\psi}^s)_{\boldsymbol{\xi}^s} + h_{\beta}^{-1}\boldsymbol{I}_{|\boldsymbol{\xi}^s|}\right)^{-1}\boldsymbol{B}(\boldsymbol{\psi}^s)_{\boldsymbol{\xi}^s}'\boldsymbol{Y},
\end{equation}
with $\boldsymbol{B}(\ddot{\boldsymbol{\psi}})$ the matrix of balances from the testing set, $\boldsymbol{B}(\boldsymbol{\psi}^s)_{\boldsymbol{\xi}^s}$ the matrix of balances selected in the $s^{th}$ MCMC iteration of the training model, and $|\boldsymbol{\xi}^s|$ the number of balances selected in $\boldsymbol{B}(\boldsymbol{\psi}^s)_{\boldsymbol{\xi}^s}$, following \cite{brown1998multivariate}. Since the $\ddot{\boldsymbol{\psi}}$ used to calculate the balances are not observed for the testing set, we estimate them as
\begin{equation}\label{psi}
\ddot{\psi}_{i,j} = \frac{\ddot{z}_{i,j} + \hat{\lambda}_{i,j}}{ \sum_{j = 1}^J \ddot{z}_{i,j} + \hat{\lambda}_{i,j}},
\end{equation}
where 
\begin{equation}\label{lambda_pred}\hat{\lambda}_{i,j} = \exp \left( \frac{1}{S} \sum_{s=1}^S \left( \alpha^s_{j} + \sum_{p=1}^P \varphi^s_{jp}\ddot{x}_{i,p}\right) \right),
\end{equation}
$\ddot{z}_i$ and $\ddot{x}_i$ represent the multivariate counts and covariates observed for the $i^{th}$ testing subject, and $\alpha^s_j$ and $ \varphi^s_{jp}$ are MCMC samples obtained from the training model. When splitting the data is impractical due to small sample sizes, leave-one-out cross-validation approximation procedures can be used, for example, following the approach proposed by  \cite{vehtari2017practical}.  This approach approximates leave-one-out (LOO) cross-validation with the expected log pointwise predictive density (epld). By using Pareto smoothed importance sampling (PSIS) for estimation, it provides a more stable estimate compared to the method of \cite{gelfand1996model}. 

\section{Case Study on Diet and the Microbiome}
We applied our joint model to dietary assessment, oral, and fecal microbiome data from an ancillary study conducted among healthy obese and lean individuals from the Houston, TX area \citep{versace2015heterogeneity}. In addition to dietary intake, physical activity, and eating behavior questionnaires, participants provided stool and oral swab samples for microbiome analysis. Participant height and weight were also measured. Adults, 21 to 55 years of age were recruited to maximize variability in usual diet/eating habits and BMI, while minimizing extraneous factors known to influence the oral and/or fecal microbiome. Individuals who used antibiotics within the past 30 days, were current smokers, had any chronic or acute condition that required exclusionary medications or dietary restrictions, reported substantial weight changes ($\pm 5$ kg) in the past 3 months, and women who were recently pregnant/lactating were excluded from the study. Approximately two-thirds of the sample were female, and 40\% were obese. Participants provided fresh stool samples using an in-home collection kit with sterile swab and no storage media between their first and second in-person visit. Study staff also collected an oral (buccal) swab sample from the participant at the in-person visit.  

Habitual dietary intake data were collected via the 134-food item National Cancer Institute Dietary History Questionnaire (DHQ) II, enabling evaluation of food groups, macronutrients, vitamins, minerals, and eight dietary supplements \citep{millen2005national,subar2001comparative}. DHQ II responses were processed via the National Cancer Institute's Diet*Calc software and initially produced 214 variables of estimated daily nutrient and food group intake. Of these, 140 variables were aggregated or excluded due to redundancy and/or low variation. The remaining 74 nutrient and food group variables were adjusted for caloric intake prior to analysis \citep{willett1998implications}. Only participants whose total energy intake was considered plausible ($800 <$ kcal $< 4200$ and $600 <$ kcal $< 3500$, for men and women, respectively) were included in this analysis. Two 24-hour dietary recalls were compared to each individual's DHQ data to assess accuracy and consistency, but not included in the current analysis.  

For microbiome assessment, stool and oral swab specimens underwent total genomic DNA extraction and 16S rDNA sequencing, as described previously \citep{hoffman2018oral,gopalakrishnan2018gut}. While highly conserved, the 16S rRNA gene is commonly used for bacterial identification due to regions of high variability \citep{li2015microbiome}. Sequencing was performed via the Illumina MiSeq platform and targeted the V4 region.  Resulting reads were processed and clustered into operational taxonomic units (OTUs) using UPARSE \citep{edgar2013uparse}  at an identity threshold of 95\%. OTUs were mapped using a V4-optimized version of the SILVA database (v.123). 
 To reduce the number of spurious relationships detected, we further limited analysis to only those OTUs identified in at least 10\% of participants. This resulted in 245 and 185 taxon for the fecal and oral samples, respectively. For consistency, only participants who provided both stool and oral swab specimens were used in this analysis, resulting in a sample size of $N=56$.

The objective of our study was to identify relations between OTUs in microbial samples and dietary covariates, while simultaneously predicting body mass index (BMI) using our proposed joint model. In two separate analyses, we modeled fecal and oral microbial samples and compared their predictive performance for BMI, controlling for age and sex by having them as fixed covariates in the model. Prior to analysis, the dietary data were standardized to mean zero and variance one. Additionally, the BMI measures were centered at the sample mean. For inference, we set hyperparameters $h_{\alpha_0} = h_{\beta} = 1$, $a_0 = b_0 = 2$, and $\sigma^2 = r^2 = 10$. Additionally, we set the hyperparameters for the beta-binomial priors to $a = a_m = 1 $, $ b = 9$, and $b_m = 4$ for both models. This corresponded to a 10\% and 20\% prior probability of inclusion for dietary factors and balances, respectively. Note, these priors were chosen since they obtained the best prediction performance in our sensitivity analysis (see end of section 3.1). The MCMC algorithm was run for 50,000 iterations, with the first 25,000 treated as burn-in and thinned every $10^{th}$ sample. In this analysis, runtimes were 16.6 and 15.8 minutes for the fecal and oral models, respectively, on a 2.5 GHz dual-core Intel Core i5 processor with 8 GB RAM. Trace plots of the log posterior distribution indicate good convergence and mixing. Covariate and balance inclusion was determined using the median model approach (i.e., MPPI $\geq$ 0.50).

\subsection{Results}
 Figures 2 and 3 show the marginal posterior probabilities of inclusion (MPPI) for dietary covariates, indexed across compositional taxa, for the model fit to the oral and fecal microbial data. 
 Figures 4 and 5 present heatmaps of the associations between dietary covariates and microbial abundances identified in the oral and fecal models, respectively. For interpretability, taxa are assigned to their likely representative bacterial genera using Basic Local Alignment Search Tool (BLAST) \citep{zhang2000greedy}.  Further details of the relations between selected pairs of taxa and dietary covariates are found in the Supplementary Material \citep{koslovsky2020}. Six balances calculated from the oral microbial sample were identified as associated with BMI, compared to 7 balances from the fecal sample. As for prediction, due to the study's relatively small sample size, we chose to compare accuracy of the results using the approach proposed by  \cite{vehtari2017practical}. We used the R package loo \citep{vehtari2016loo}, which requires the pointwise log-likelihood, $f(y_i|\xi^s, \psi^s_{i})$, for each subject $i = 1, \dots , N$ calculated at each MCMC iteration $s = 1, \dots, S$, and produces an estimated $\widehat{\mbox{epld}}$ value, with larger values implying a superior model. In our analysis, the models fit with the oral and fecal data provided similar results  ($\widehat{\mbox{epld}}_{\mbox{\tiny ORAL}} = -200.2$ versus $\widehat{\mbox{epld}}_{\mbox{\tiny FECAL}} = -201.6$, respectively).

\begin{figure}
  \caption{Marginal posterior probabilities of inclusion for dietary covariates indexed across compositional taxa using the oral microbial data. Dashed line represents the median model threshold (0.50).}
  \centering
\includegraphics[scale = 0.5]{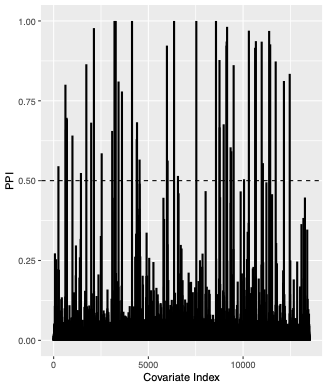}
\end{figure}

\begin{figure}
  \caption{Marginal posterior probabilities of inclusion for dietary covariates indexed across compositional taxa using the fecal microbial data. Dashed line represents the median model threshold (0.50).}
  \centering
\includegraphics[scale = 0.5]{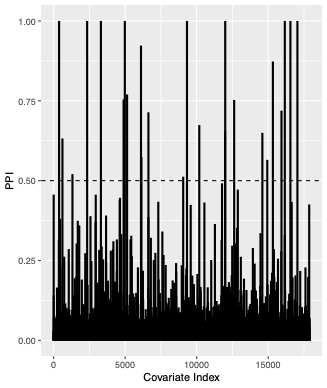}
\end{figure}

Causal links between diet, the gut microbiome, and BMI/obesity are becoming clearer \citep{ridaura2013gut,maruvada2017human,turnbaugh2017microbes}. Microbiota play a key role in the extraction, absorption and storage of energy from dietary intake. Some of the most compelling findings are for ``Western-style'' dietary patterns, which are typically characterized by low intake of fiber-rich plant foods and high intake of meat and added sugars, leading to a microbiome that is generally more hospitable and supportive of the development of obesity and other chronic diseases \citep{turnbaugh2017microbes, valdes2018role}. Interestingly across both the oral and fecal microbiome, we observed several associations with ``Western-style'' dietary factors and their counterparts, e.g., different nutrient rich and prebiotic vegetable food groups, as well as various B vitamins (those largely found in animal sources) and antioxidant nutrients derived from both dietary intake and supplement use.

\begin{sidewaysfigure}\label{graphical_oral}
  \caption{ Estimated regression coefficients for corresponding dietary factors and oral microbial abundance associations, grouped by BLAST assignment, identified using the proposed joint model.  }
  \centering
\includegraphics[scale = 0.6]{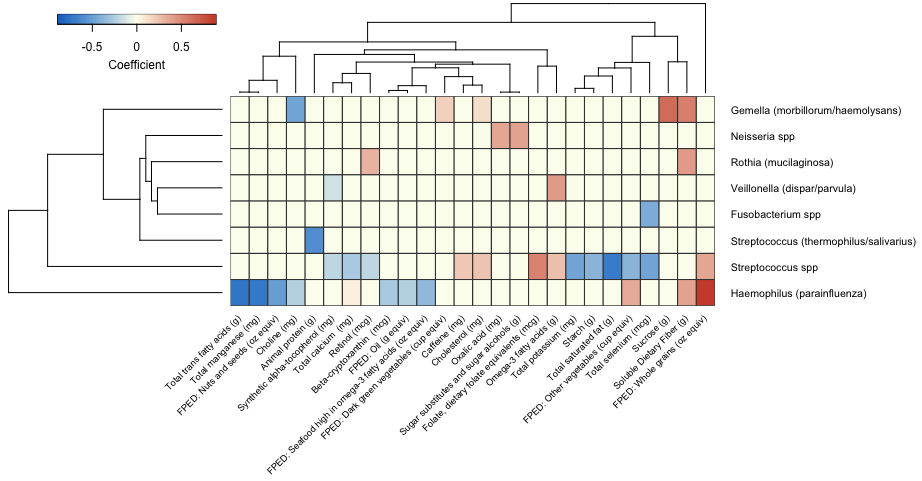}
\end{sidewaysfigure}

\begin{sidewaysfigure}\label{graphical_fecal}
  \caption{  Estimated regression coefficients for corresponding dietary factors and fecal microbial abundance associations, grouped by BLAST assignment, identified using the proposed joint model.}  
  \centering
\includegraphics[scale = 0.6]{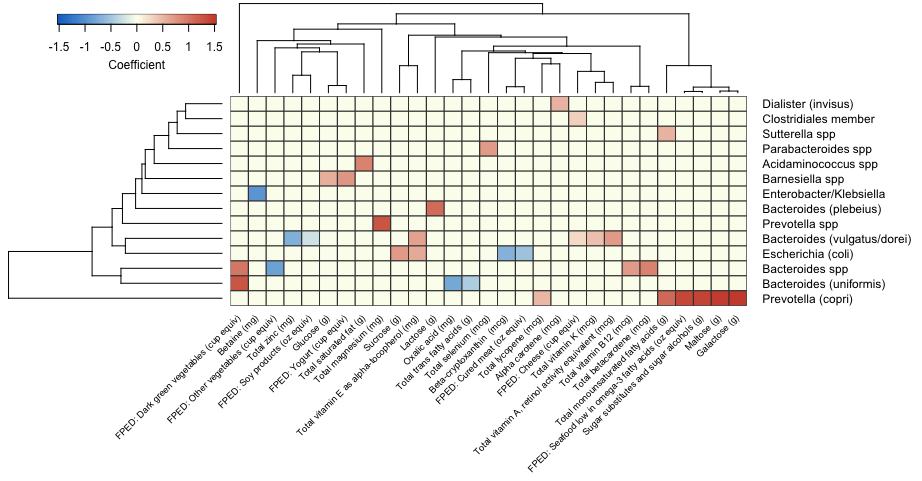}
\end{sidewaysfigure}

Looking at the results on the fecal microbiome, we observe several dietary relationships with Bacteroides, including lactose and consumption of dark green vegetables. Several Bacteroides species (a common and abundant genus within Bacteroidetes phylum) and their interactions with diet have been implicated in obesity \citep{kovatcheva2015dietary,david2014diet,wu2011linking}. Bacteroides have a broad capacity to use diverse types of carbohydrates or dietary polysaccharides, which include glucose, sucrose, and starch, for energy; and can ``step up'' when dietary fiber intake is low, tapping into other sources of energy for the gut \citep{sonnenburg2010specificity,gurry2018predictability,marcobal2011bacteroides}. In a recent and similarly conducted epidemiologic study of healthy adults, low fiber intake was associated with higher Bacteroides uniformis \citep{lin2018association}. As with Bacteroides, Escherichia also metabolizes carbohydrates for energy and has been associated with intestinal inflammation in animal models fed a Western (high fat-high sugar) diet \citep{agus2016western,martinez2014western}, consistent with the Escherichia-sucrose association observed in this analysis.  We additionally found a number of associations linked to Prevotella, including maltose, galactose, and sugar substitutes and alcohols - commonly found in snack foods.  Greater levels of Prevotellaceae have been observed in obese individuals \citep{zhang2009human}, and Prevotella copri specifically has been found in higher abundance among overweight and obese type 2 diabetics \citep{leite2017detection}.

Similar to the gut microbiome, the oral microbiome may also be shaped by dietary habits \citep{hansen2018impact,ercolini2015imbalance,kato2017nutritional,peters2018association,fan2018drinking}. Differences in the diversity and abundance of oral bacteria between overweight/obese and healthy weight individuals have now been documented in several studies \citep{goodson2009obesity,zeigler2012microbiota,haffajee2009relation}.  In particular, \cite{yang2019oral} found increased oral Gemella and Streptococcus oligofermentans  among obese persons in a large ($n>1,500$) cohort study.  Gemellaceae and Streptococcaceae were also more abundant in obese subjects whose saliva suppressed aromatic compounds from wine, and the authors note that altered sensory responses may result in greater food intake \citep{piombino2014saliva}.  While Gemella was linked to both sucrose and cholesterol intake in our study, Streptococcus was negatively associated with key Western diet components: namely, starch, animal protein, and total saturated fat.  This is likely explained by species-level variation, which cannot be definitely determined by 16S sequencing; but it is important to note that Streptococcus members are the most abundant bacteria of the mouth \citep{huttenhower2012structure}. Taken overall, the current evidence suggests that the microbiome may be a reflection of obesity (or leanness), as well as a cause of it, largely via diet-microbiome interactions \citep{ridaura2013gut,komaroff2017microbiome}.

While there are no methods available for direct comparison to our joint model, we compared the results of our analysis to two, two-step approaches that first select dietary covariates associated with fecal and oral multivariate count data and then perform variable selection on balances, constructed using estimated count probabilities, for prediction of BMI. In the first step, we used a recently proposed Bayesian variable selection method for Dirichlet-multinomial regression models (DM-BVS) \citep{wadsworth2017integrative}  and a penalized approach introduced by \cite{chen2013variable} (CL).   For the CL approach, the group penalty was set to 20\%, and the model with the lowest Bayesian information criterion was used for inference \citep{schwarz1978estimating}.  In the second step, we fit a multiple linear regression model and performed variable selection on the balances calculated using the estimated $\boldsymbol{\psi}$ obtained in step one. The method for obtaining estimates of $\boldsymbol{\psi}$ differed across models, as explained in the section 4. We applied Bayesian variable selection for the DM-BVS approach and the lasso for the CL approach \citep{george1997approaches,tibshirani1996regression}.  We refer to the Bayesian and penalized two-step approaches as DMLM-Bayes and DMLM-Pen, respectively. Both methods were compared in regards to their selection of covariates as well as their model fit.

  For the oral microbial data, the DMLM-Pen and DMLM-Bayes approaches selected 61 and 45 covariate-taxon relations, respectively (see Supplementary tables S3 and S4 \citep{koslovsky2020}). Using the DMLM-Pen approach, only two relations were also found using our joint model. However using the DMLM-Bayes approach, seven relations were also found using our method. For the fecal microbial data, the DMLM-Pen and DMLM-Bayes approaches selected 15 and 23 covariate-taxon relations, respectively (see Supplementary Tables S5 and S6 \citep{koslovsky2020}). Similarly using the DMLM-Pen approach, only three relations were also found using our joint model. However using the DMLM-Bayes approach, eleven relations were also found using our method. Additionally to assess model fit,  the mean squared error (MSE) for the joint model applied to the fecal data was 11.45, compared to 2874.34 and 27.98 for the DMLM-Pen and DMLM-Bayes approaches, respectively. Similarly, the MSE for the oral data was 90.94 with the joint model and 2993.28 and 126.16 with the DMLM-Pen and DMLM-Bayes approaches, respectively. While all models fit the fecal data better, our joint model demonstrated superior model fit for both the fecal and oral data.

We performed a sensitivity analysis to assess the sensitivity of the results produced by the joint model to prior specification. Specifically, we investigated differences in the selection and prediction results with  $a = a_m = 1 $ and $ b = b_m = \{1,4,9\}$  as well as $a_0 = 2$ and $b_0 = \{2,4,16,256\}$   for both the fecal and oral models separately. As expected, the number of covariates and balances selected in the model increased as the prior probability of inclusion increased. Similar to the sensitivity analysis on simulated data (section 4), prediction performance diminished as $b_0$ increased.  

\section{Simulation Study}

In this section, we evaluate the selection performance and predictive ability of our proposed joint model using simulated data. Performance is compared to the two two-step approaches presented in the case study. The method for obtaining estimates of $\boldsymbol{\psi}$ differs across models, as explained below. 

We simulated $N = 50$ subjects with $P = 50$ covariates and  $J = 150$  compositional taxa. Covariates $\boldsymbol{x}$ were simulated from a $N_P(\boldsymbol{0},\Sigma)$, where $\Sigma_{i,j} = \omega^{|i-j|}$ and $ \omega = 0.4$. In each of the replicate datasets, we randomly selected 10 of the 7,500 covariate-taxon combinations to be associated with the compositional data. Corresponding regression coefficients  $\varphi $ were randomly sampled from  $ \pm[0.75,1.25]$. Intercept terms $\boldsymbol{\alpha}$ were simulated from a uniform$[-2.3,2.3]$. The multivariate count data ${\bf Z}$ were sampled from a Multinomial($\dot{z}_i, \psi_i^*$), where $\dot{z}_i \sim \mbox{uniform[2,500, 7,500]}$ and $\psi_i^*  \sim \mbox{Dirichlet}(\boldsymbol{\gamma}_i^*)$, where $\boldsymbol{\gamma}_i^* = (\gamma_{i,1}^*,\gamma_{i,2}^*,\dots,\gamma_{i,J}^*)$. Each $\gamma_{i,j}^* = \frac{\gamma_{i,j}}{\sum_{j=1}^J \gamma_{i,j}}\frac{1-d}{d}$, $j = 1,\dots,J$, where $\gamma_{i,j}$ was determined using equation \eqref{lambda}, and $d$ serves as an overdispersion parameter which was set at $0.01$, similar to \cite{wadsworth2017integrative, chen2013variable}. Thus, the data generating model differs from our model assumptions. We used a pseudovalue of $ 6.67 \times 10^{-5}$ to replace zero values of $\psi_{i,j}$, which corresponds to the maximum roundoff error, $0.5$, divided by the maximum possible value of $\dot{z}_i$, 7,500.  This is done to prevent taking the log of zero when calculating balances.  We then generated the response data as $y_i = \alpha_0 + \boldsymbol{B}^*(\psi_i^*)'\boldsymbol{\beta} + \epsilon_i$, where $\alpha_0 = 0$, $\boldsymbol{\beta}$ is a $J-1$-dimensional vector of regression coefficients, $\boldsymbol{B}^*(\psi_i^*)$ are the balances calculated using sequential binary separation, and $\epsilon_i \sim N(0,1)$. Of the $J-1$ regression coefficients, 5 were randomly sampled from $ \pm[1.25,1.75]$ and the rest were set equal to zero. 

When running the MCMC, we set hyperparameters $h_{\alpha_0}=h_{\beta}=1$, as well as $a=1$, and $b=\{9,99,999\}$, representing a prior expectation of 10\%, 1\%, and 0.1\% of the total number of covariates included in the model. For balance selection, $a_m$ and $b_m$ were set similarly. Before analysis, $\boldsymbol{y}$ was mean-centered, and covariates and balances were standardized to mean zero and variance one. Note that in our joint model, balances are standardized at each MCMC iteration since they are recalculated using the current iteration's $\boldsymbol{\psi}_i$. Simulations were run for 20,000 iterations and thinned to every $10^{th}$ iteration. This resulted in 2,000 iterations, of which the first 1,000 iterations were treated as burn-in, and the remaining 1,000 used for inference. Each run was initiated with a random 1\% of the 7500 covariate-taxon combinations' and 5\% of the 149 balances' inclusion indicators active. Covariates and balances were determined to be associated with the compositional and response data, respectively, if their MPPI $\geq$ 0.50 \citep{barbieri2004optimal}. Results we report below were obtained by averaging over 30 replicated datasets.

For variable selection, all methods were assessed on the basis of sensitivity (1 - false negative rate), specificity (1 - false positive rate), and Matthew's correlation coefficient (MCC) (a measure of overall selection accuracy). These are defined as: 
$$\mbox{Sensitivity} = \frac{TP}{FN + TP}$$
$$\mbox{Specificity} = \frac{TN}{FP + TN}$$
$$MCC = \frac{TP \times TN - FP \times FN}{\sqrt{(TP +FP)(TP+FN)(TN+FP)(TN+FN)}} ,$$
where TN, TP, FN, and FP represent the true negatives, true positives, false negatives, and false positives, respectively. To assess prediction performance, we trained the models on the 50 samples used for variable selection and tested the models on an additional 50 samples generated similarly. Prediction accuracy was assessed with the predicted mean squared error (PMSE), defined as $\sum_{i = 1}^{50} (\mathcal{Y}_{i} - \hat{\mathcal{Y}}_i )^2$, where $\mathcal{Y}_i$ is from the testing set and $\hat{\mathcal{Y}}_i$ is its predicted value. To obtain predictions of the outcomes with our joint model, we followed equation \eqref{predictY}. Model fit was assessed with mean squared error (MSE), defined as $\sum_{i = 1}^{50} (Y_{i} - \hat{Y}_i)^2$, where $Y_i$ is from the training set. To obtain an estimate of the outcomes, $\hat{Y_i}$, we followed the approach used to calculate the PMSE, replacing $\boldsymbol{\mathcal{Y}}$ and $ \boldsymbol{B}(\ddot{\boldsymbol{\psi}})$, with $\boldsymbol{Y}$ and $\boldsymbol{B}(\boldsymbol{\psi}^s)_{\boldsymbol{\xi}^s}$, respectively. Estimates for the DMLM-Bayes approach were obtained similarly, with the exception that the average of the $S$ MCMC samples of $\boldsymbol{\psi}$ from the first step were used to construct $\boldsymbol{B}( \boldsymbol{\psi} )$ in the second. For the DMLM-Pen approach, the testing balances are estimated using a similar approach as above, replacing the average of the MCMC samples in equations \eqref{psi} and \eqref{lambda_pred} with the CL model estimates.

\subsection{Results}

Tables \ref{Simu:covariate} and \ref{Simu:balance} report results for the proposed joint model (JM), the two-step Bayesian approach (DMLM-Bayes) and the two-step penalized approach (DMLM-Pen) in terms of sensitivity, specificity, MCC, MSE, and PMSE averaged over 30 simulations with standard errors in parentheses. For the Bayesian models, results are assessed over various beta-binomial priors for a covariate's probability of inclusion. Note that JM and DMLM-Bayes have similar performance for covariate selection since the underlying models are the same. Thus, we only compare to the DMLM-Pen approach in Table \ref{Simu:covariate}. For the selection of covariates associated with the multivariate count data, both of the models showed high specificity (Table \ref{Simu:covariate}). Note that models may have selected unassociated terms and still obtained a specificity of 1.00 due to rounding.  However, the JM outperformed the DMLM-Pen approach in terms of sensitivity and MCC. The JM with hyperparameters $a= 1 \mbox{ and }b=99$ performed the best overall. As expected, the number of covariates selected was reduced as the mean of the inclusion prior decreased for the Bayesian approach. The  DMLM-Pen approach selected the most covariates on average, leading it to have the lowest MCC overall. For the selection of balances associated with the continuous response, we found similar results for all of the models in terms of specificity (Table \ref{Simu:balance}). For the Bayesian methods, the number of selected balances, as well as the sensitivity and MCC went down as the prior probability of inclusion decreased. Overall, the DMLM-Bayes model with hyperparameters $a=1$ and $b=9$ performed the best, with a similar performance achieved by the JM. We observed the worst performance in terms of sensitivity, specificity, and MCC for the DMLM-Pen approach, as a result of the poor estimation of $\boldsymbol{\psi}$ from the DM portion of the model. Trace plots of the log posterior showed good mixing, and no observed trends in the plots after burn-in suggested model convergence across the simulations. In simulation results not shown, all of the methods maintained extremely high specificity for the null model which contained no true relations between covariates and the compositional data. Also, we found that selection performance was not sensitive to replacement pseudovalues for $\psi_{i,j} = 0$ during sampling. 

\begin{table*} 
{ \footnotesize
\centering
\label{covariate}
\caption{Covariate selection simulation results for the proposed joint model (JM) and two-step penalized DMLM-Pen approach in terms of sensitivity (Sens.), specificity (Spec.), and Matthew's correlation coefficient (MCC) averaged over 30 simulations with standard deviations in parentheses. For Bayesian models, results are assessed over various beta-binomial priors for covariates' probability of inclusion, $\boldsymbol{\zeta}$. \label{Simu:covariate}}
\setlength\tabcolsep{1.5pt} 
\begin{tabular}{ccccccccccccccc}
\hline
 & \multirow{2}{*}{\textbf{\begin{tabular}[c]{@{}c@{}}Selection \\ Prior\end{tabular}}} &  & \multicolumn{4}{c}{\textbf{Covariates}} \\ \cline{4-7} 
 &  &  & Selected & Sens. & Spec. & MCC  \\ \hline
\textbf{} & $a=1$, $b=9$ &  & 12.93 (4.15) & 0.83 (0.16) & 1.00 (0.00) & 0.74 (0.13)   \\
\textbf{JM} & $a=1$, $b=99$ &  & 7.47 (2.01) & 0.71 (0.18) & 1.00 (0.00) & 0.82 (0.12)   \\
\textbf{} &  $a=1$, $b=999$ &  & 6.20 (2.34) & 0.61 (0.24) & 1.00 (0.00) & 0.78 (0.16)  \\
\textbf{DMLM-Pen} & - &  & 25.00 (21.67) & 0.30 (0.19) & 1.00 (0.00) & 0.20 (0.06)   \\ \hline
\end{tabular}
}

\end{table*}

\begin{table*}[bp!]
{ \setlength{\tabcolsep}{3.5pt}
{ \footnotesize
\centering 
\caption{Balance selection simulation results for the proposed joint model (JM), the two-step Bayesian approach (DMLM-Bayes), and the two-step penalized approach (DMLM-Pen) in terms of sensitivity (Sens.), specificity (Spec.), and Matthew's correlation coefficient (MCC) averaged over 30 simulations with standard deviations in parentheses. For Bayesian models, results are assessed over various beta-binomial priors for balances' probability of inclusion, $\boldsymbol{\xi}$.\label{Simu:balance}}

\begin{tabular}{cccccccccc}
\hline
 & \multirow{2}{*}{\textbf{\begin{tabular}[c]{@{}c@{}}Selection \\ Prior\end{tabular}}} &   & \multicolumn{4}{c}{\textbf{Balances }}  \\
 \cline{4-7} 
 &  &  & Selected & Sens. & Spec. & MCC   \\ \hline
\textbf{} & $a=1$, $b=9$ & & 9.30 (0.99) & 0.92 (0.09) & 1.00 (0.00) & 0.94 (0.06)   \\
\textbf{JM} & $a=1$, $b=99$ & & 3.87 (2.61)  & 0.38 (0.27) & 1.00 (0.00) & 0.55 (0.23)   \\
\textbf{} & $a=1$, $b=999$ & & 0.80 (1.10) & 0.08 (0.11) & 1.00 (0.00) & 0.38 (0.11)   \\
\textbf{} & $a=1$, $b=9$ & & 10.33 (0.76) & 0.97 (0.06) & 1.00 (0.01) & 0.95 (0.07)  \\
\textbf{DMLM-Bayes} & $a=1$, $b=99$ &  & 6.87 (3.22) & 0.87 (0.34) & 1.00 (0.00) & 0.79  (0.25) \\
\textbf{} & $a=1$, $b=999$ & & 1.23 (1.52) & 0.12 (0.15) & 1.00 (0.00) & 0.42 (0.15)   \\
\textbf{DMLM-Pen} & - &  & 1.17 (1.46) & 0.04 (0.09) & 0.99 (0.01) & 0.11 (0.19)  \\ \hline
\end{tabular}
}}
\end{table*}

In terms of model fit, the DMLM-Bayes two-step approach with hyperparameters $a=1$ and $b=9$ had the smallest MSE on average, as expected given its balance selection performance (Table \ref{Simu:prediction}). For both Bayesian approaches, the average MSE increased with more informative priors. This is mainly due to diminished sensitivity for both covariates and balances. Our joint model with weakly-informative priors had the lowest PMSE on average, closely followed by the DMLM-Bayes approach with similar prior specification. Despite its improved prediction performance, the JM had relatively higher PMSE standard deviations compared to the DMLM-Bayes approach, as hypothesized.  The DMLM-Pen approach had the largest MSE and PMSE overall, reflecting its relatively poor selection performance for both covariates and balances.

\begin{table}[bp!] 
{ \setlength{\tabcolsep}{6pt}
{ \footnotesize
\centering
\caption{Simulation results for the proposed joint model (JM), the two-step Bayesian approach (DMLM-Bayes), and the two-step penalized approach (DMLM-Pen) in terms of mean squared error (MSE) and prediction mean squared error (PMSE) averaged over 30 simulations with standard deviation in parentheses. For Bayesian models, results are assessed over various beta-binomial priors for a covariate's probability of inclusion.\label{Simu:prediction}}
\begin{tabular}{ccccc}
\hline
&  \textbf{Selection  Prior} &   &  \textbf{MSE} & \textbf{PMSE}  \\
\hline 
\textbf{} & $a=1$, $b=9$ &  & 101.28 (31.74) & 563.54 (226.62) \\
\textbf{JM} & $a=1$, $b=99$ &   & 1250.13 (734.24) & 1953.09 (1019.78) \\
\textbf{} & $a=1$, $b=999$ &    & 3214.91 (974.96) & 3494.93 (913.75) \\
\textbf{} & $a=1$, $b=9$ &   & 67.22 (16.93) & 785.63 (327.73) \\
\textbf{DMLM-Bayes} & $a=1$, $b=99$ &   & 509.03 (580.00) & 2527.30 (1220.18) \\
\textbf{} & $a=1$, $b=999$ &   & 2710.42 (1134.56) & 3562.42 (819.85) \\
\textbf{DMLM-Pen} & - &   & 3521.97 (863.37) & 4267.24 (1206.15) \\ \hline
\end{tabular}
}}
\end{table}

\subsection{Sensitivity Analysis}

\begin{table}[bp!]
{ \setlength{\tabcolsep}{4pt}
\centering
\caption{Results of sensitivity analysis for hyperparameter $b_0$ in Inverse-gamma prior for total number of selected covariates across taxa and balances (\#), sensitivity (Sens.), specificity (Spec.), Matthew's correlation coefficient (MCC), and mean squared error (MSE).\label{Sensitivity} }

\begin{tabular}{cccc cccc cccc cc}
\hline
\multirow{2}{*}{$b$} & \multirow{2}{*}{$b_0$} &  \multicolumn{4}{c}{\textbf{Covariates}}                    &  & \multicolumn{4}{c}{\textbf{Balances}}                        &  & \multirow{2}{*}{\textbf{MSE}} & \multirow{2}{*}{\textbf{PMSE}} \\ \cline{3-6} \cline{8-11}
                       &  & \# & Sens. & Spec. & MCC &  & \# & Sens.  &  Spec. &  MCC &  &                               \\ \hline
\multirow{4}{*}{$9$} & 1  & 10                & 0.90            & 1.00              & 0.90         &  & 8                 & 0.80              & 1.00           & 0.89         &  & 112.76 & 619.39                        \\
& 2  & 10                & 0.90            & 1.00              & 0.90         &  & 8                 & 0.80              & 1.00           & 0.89         &  & 111.84 & 616.63                        \\
& 4  & 10                & 0.90            & 1.00              & 0.90         &  & 9                 & 0.80              & 0.99           & 0.83         &  & 117.54 & 680.12                      \\
& 8   & 10                 & 0.90            & 1.00              & 0.90         &  & 5                 & 0.50              & 1.00             & 0.69            &  & 510.74 & 1063.09                  \\
 \hline
 \multirow{4}{*}{$99$} & 1  & 10                 & 1.00            & 1.00              & 1.00        &  & 7                 & 0.70              & 1.00           & 0.83         &  & 340.21 & 691.27                     \\
 & 2  & 10                 & 1.00            & 1.00              & 1.00        &  & 7                 & 0.70              & 1.00           & 0.83         &  & 340.25 & 694.94                                 \\
& 4 & 10                 & 1.00            & 1.00              & 1.00        &  & 6                 & 0.60              & 1.00           & 0.76         &  & 428.12 & 768.15                    \\       
& 8  & 10                 & 1.00          & 1.00              & 1.00        &  & 1                 & 0.10              & 1.00           & 0.31         &  & 2073.22 & 2125.34                     \\
 \hline
 \multirow{4}{*}{$999$} & 1  & 9                & 0.80            & 1.00              & 0.84         &  & 1                 & 0.10              & 1.00           & 0.31         &  & 2646.49 & 2427.75                        \\
& 2   & 9                 & 0.80            & 1.00              & 0.84         &  & 1                 & 0.10              & 1.00           & 0.31         &  & 2646.49 & 2427.75                        \\
& 4 & 9                 & 0.80            & 1.00              & 0.84         &  & 1                 & 0.10              & 1.00           & 0.31         &  & 2882.92 & 2568.60                       \\
& 8  & 9                 & 0.80            & 1.00              & 0.84         &  & 1                 & 0.10              & 1.00           & 0.31         &  & 3156.92 & 2735.28                      \\
 \hline
\end{tabular}}
\end{table}

We investigated the model sensitivity to specification of hyperparameters $b_0$, $a$, and $b$. In each of the sensitivity analyses, replicate data generated from the model defined in the simulation section were used. We evaluated the number of covariates selected, sensitivity, specificity, MCC, MSE, and PMSE for the scale parameter in the Inverse-gamma prior for the error variance, $b_0$, at values in the set $\left\lbrace 1, 2, 4, 8  \right\rbrace$ (on $\log_2$ scale), holding $a_0 = 2$. With this parameterization, $b_0$ can interpreted as the expectation of $\sigma^2$. Additionally, we assessed the model's sensitivity to different beta-binomial priors for the inclusion indicators. Specifically, we used a weakly ($a=1\mbox{,  }b=9$), moderately ($a=1\mbox{,  }b=99$), and highly ($a=1\mbox{,  }b=999$) informative prior, with $E[\zeta_{jp}] = 0.1, 0.01$ and $0.001$, respectively. 

To assess the sensitivity of the model to the specification of the Inverse-gamma prior for the random error $\sigma^2$, we set $a_0=2$ and fit the model across a range of $b_0$. The results of our sensitivity analysis are presented in Table \ref{Sensitivity}. As expected selection performance for the covariates associated with taxa probabilities were unaffected by $b_0$. However, we observed a negative relation between $b_0$ and the number of balances selected, as well as balance sensitivity, specificity, and MCC. As a result, we observed a positive relation between $b_0$ and MSE/PMSE. Additionally, the number of selected covariates and balances decreased with the expected prior probability of inclusion. 

\section{Discussion}
In this work, we have presented a Bayesian model for jointly identifying dietary covariates that are associated with microbial data and predicting a continuous, phenotypic response using a set of balances constructed from the estimated compositional taxa probabilities. Our approach induces sparsity on both balances and covariates while incorporating the structure of the multivariate count data. In our application, we found numerous associations between microbial taxa and dietary factors that may lead to a microbiome that is generally more hospitable to the development of chronic diseases, such as obesity. Additionally, we observed similar prediction performance of BMI for fecal and oral microbiome data. Through simulation, we have demonstrated the benefits of jointly modeling these data in terms of covariate selection performance and prediction accuracy. Additionally, we show how the Bayesian two-step approach had lower prediction accuracy and may underestimate prediction uncertainty by treating the compositional count data as fixed. In clinical applications, this may result in overconfident prediction estimates of the phenotypic response, which may promote the implementation of ineffective treatments or intervention strategies. While designed to study microbial abundance data, our method can handle any research setting in which multivariate count data may mediate the relation between a set of risk factors and a continuous response. Thus our proposed model is agnostic to the sequencing approach used to quantify microbial samples.

 Our model provides an integrated analysis of the relations between behavioral, microbial, and phenotypic measures collected on a cohort of healthy obese and lean individuals. Given the complexity of the model, full validation of clinical results requires the availability of data collected on dietary covariates, fecal and/or oral microbiome samples, BMI, as well as potential confounders (i.e., age and sex). However, the conditional independence structure implied by the joint model allows researchers to validate key aspects separately. For example, the selected associations between individual dietary factors and microbial counts can be directly compared to other studies investigating these relations. In these settings, reproducibility is primarily challenged by vast heterogeneity in microbial abundances found across individuals and populations \citep{huttenhower2012structure,takeshita2014distinct,li2014comparative,falony2016population} as well as study design issues including differences in food frequency questionnaires \citep{bowyer2018use}.  Another key aspect of our model is its ability to accommodate taxa heterogeneity when predicting phenotypic responses. While our case study was not large enough to justify out-of-sample validation, larger follow-up studies could assess predictive performance using the cross-validation approach described at the end of section 2.1. 

While our approach provides unique insights into the relation between modulating factors and phenotypic outcomes via microbial composition samples, it currently lacks the ability to accommodate repeated measures data collected in longitudinal studies. The ability to model both fixed and random effects would allow researchers to investigate how the relations between diet, microbiome, and BMI vary over time and across subjects. Additionally, structural information on phylogenetic trees could be incorporated into the multinomial distribution used to model the relation between covariates and the multivariate count data using a Dirichlet-tree multinomial model, which permits both positive and negative correlation structures among the count data \citep{tang2018phylogenetic,wang2017dirichlet}. Also, our approach is developed for exploratory data analysis settings designed to generate hypotheses regarding the relations among covariates, compositional data, and a response. In more confirmatory settings, researchers may aim to assess treatment effects on microbial composition as well as the phenotypic response, while controlling for a set of possible confounders. Oftentimes, the appropriate subset of confounders to control for may be unknown, and the space to search through is large compared to the number of observations. In this setting, our approach could be extended to search the pool of potential confounders in human microbiome studies following the methods proposed in \cite{antonelli2017high}. In this analysis, we construct balances using binary sequential separation and focus our inference on prediction, not explanation. Future studies could incorporate biological information when constructing balances, similar to \cite{silverman2017phylogenetic,morton2017balance,washburne2017phylogenetic}, and additionally investigate the relations between balances and phenotypic responses. Lastly, our approach is presented for continuous outcomes, but discrete, as well as survival outcomes are often encountered in biomedical settings. To handle discrete phenotypic outcomes, such as disease onset, the joint model could easily be adjusted using data augmentation approaches \citep{albert1993bayesian,polson2013bayesian}.




\section*{Acknowledgements}
Matthew Koslovsky is supported by NSF via the Research Training Group award DMS-1547433. Data utilized from the diet and microbiome study was supported by a grant to Carrie Daniel from The University of Texas MD Anderson Cancer Center Duncan Family Institute for Cancer Prevention and Risk Assessment. Carrie Daniel's efforts on this project are further supported by the NIH/NCI Cancer Center Support Grant P30 CA016672.

\begin{supplement}  
\stitle{Supplemental Code and Tutorial}
\slink[doi]{COMPLETED BY THE TYPESETTER}
\sdescription{To help researchers use our approach, we provide R code and an accompanying tutorial applying our approach to simulated data. To enhance the performance of our approach, we integrated C++ into our source code using Rcpp and RcppArmadillo \citep{eddelbuettel2014rcpparmadillo,eddelbuettel2011rcpp}. The code developed for this manuscript, simulated data, and a worked example are publicly available on GitHub. \url{https://github.com/mkoslovsky/DMLMbvs}}
\end{supplement}

\begin{supplement}  
\stitle{Supplemental Simulations and Results}
\slink[doi]{COMPLETED BY THE TYPESETTER}
\sdatatype{.pdf}
\sdescription{In this document, we provide an additional simulation study demonstrating the model's invariance to balance specification and how balance sparsity can improve prediction performance, as well as additional tables and figures containing results of our case study analysis.}
\end{supplement}
 
\bibliographystyle{imsart-nameyear}
\bibliography{microB_joint}

\end{document}